\documentclass[12pt,a4]{article}
\usepackage{amsmath}
\usepackage{bm}
\usepackage{amsfonts}
\usepackage{amssymb}
\usepackage{graphics}
\usepackage[normal]{caption2}
\usepackage{subfigure}
\usepackage{rotating}
\usepackage{citesort}
\def\be{\begin{equation}}
\def\ee{\end{equation}}
\def\ba{\begin{array}}
\def\ea{\end{array}}
\def\bea{\begin{eqnarray}}
\def\eea{\end{eqnarray}}

\begin{document}
\baselineskip 20pt \setlength\tabcolsep{2.5mm}
\renewcommand\arraystretch{1.5}
\setlength{\abovecaptionskip}{0.1cm}
\setlength{\belowcaptionskip}{0.5cm}
%%%%%%%%%%%%%%%%%%%%%%%%%%%%%%%%%%%%%%%
\pagestyle{empty}
\newpage
\pagestyle{plain} \setcounter{page}{1} \setcounter{lofdepth}{2}
\begin{center} {\large\bf Isospin effects in the disappearance of flow as a function of colliding geometry}\\
\vspace*{0.4cm}
{\bf Sakshi Gautam$^a$}, {\bf Aman D. Sood$^b$}\footnote{Email:~amandsood@gmail.com}, {\bf Rajeev K. Puri$^a$}, {\bf and J. Aichelin$^b$}\\
$^a${\it  Department of Physics, Panjab University, Chandigarh
-160 014, India.\\} $^b${\it  SUBATECH, Laboratoire de Physique
Subatomique et des Technologies Associ\'{e}es, Universit\'{e} de
Nantes - IN2P3/CNRS - EMN 4 rue Alfred Kastler, F-44072 Nantes,
France.\\}
\end{center}
We study the effect of isospin degree of freedom on the balance energy
(E$_{bal}$) as well as its mass dependence throughout the mass
range 48-270 for two sets of isobaric systems with N/Z = 1 and 1.4 at different colliding geometries ranging from central to peripheral ones. Our
findings reveal the dominance of Coulomb repulsion in isospin
effects on E$_{bal}$ as well as its mass dependence throughout the
range of the colliding geometry. Our results also indicate that the effect
of symmetry energy and nucleon-nucleon cross section on E$_{bal}$ is uniform throughout the mass range and throughout the
colliding geometry. We also present the counter balancing of nucleon-nucleon collisions and mean field by reducing the Coulomb and the counter
balancing of Coulomb and mean filed by removing the nucleon-nucleon collisions.

%%%%%%%%%%%%%%%%%%%%%%%%%%%%%%%%%5

\newpage
\baselineskip 20pt
\section{Introduction}
 With the availability of radioactive ion beam (RIB) facilities at Cooler Storage Ring (CSR)
(China) \cite{rib1}, the GSI Facility for Antiproton and Ion beam
Research (FAIR) \cite{rib2}, RIB facility at Rikagaku Kenyusho
(RIKEN) in Japan \cite{rib3}, GANIL in France \cite{rib4}, and the
upcoming facility for RIB at Michigan State University \cite{rib5}
one has a possibility to study the properties of nuclear
matter under the extreme conditions of isospin asymmetry. Heavy-ion
collisions induced by the neutron rich matter provide a unique
opportunity to explore the isospin dependence of in-medium nuclear
interactions, since isospin degree of freedom plays an important
role in heavy-ion collisions through both the nuclear matter equation of
state (EOS) as well as via in-medium nucleon-nucleon (nn) cross section.
\par
 After about three decades of intensive efforts in both
nuclear experiments and theoretical calculations, equation of state for isospin
symmetric matter is now relatively well understood \cite{lirpt}. The effect of isospin degree of
freedom on the collective transverse in-plane flow as well as on its disappearance \cite{krof89} (there exists a particular
incident energy called \emph{balance energy} (E$_{bal}$) or \emph{energy of vanishing flow} (EVF) at which transverse in-plane flow
disappears) has been reported in the literature \cite{li96,pak97,daff98}, where it was found
 that neutron-rich systems have higher  E$_{bal}$ compared to neutron-deficient
  systems at all colliding geometries varying from central to peripheral
ones. The effect of isospin degree of freedom on  E$_{bal}$ was found to be much more pronounced at peripheral colliding geometries
 compared to central ones. As reported in the literature, the isospin
dependence of collective flow as well as its disappearance has been explained as
a competition among various reaction mechanisms, such as
nucleon-nucleon collisions, symmetry energy, surface property of
the colliding nuclei, and Coulomb force. The relative importance
among these mechanisms is not yet clear \cite{li96}. In recent study, we \cite{gaum10}
confronted theoretical calculations (using isospin-dependent quantum molecular
dynamics (IQMD) model \cite{hart98}) with the data at all colliding geometries and
were able to reproduce the data within 5\% on average at all colliding geometries. Motivated by the good agreement
of the calculations with data, two of us \cite{gaum210} (in order to explore the relative
importance among above mentioned reaction mechanisms in isospin effects on the
 E$_{bal}$) calculated the  E$_{bal}$ throughout the mass range for two sets
 of isotopic systems with N/Z =1.16 and 1.33. The isotopic pairs were
 chosen so that the effect of the Coulomb repulsion is the same for a given pair.
 The choice of the above N/Z was taken because the percentage difference $\Delta$ N/Z$ (\%)$ = $\frac{(N/Z)^{1.33}-(N/Z)^{1.16}}{(N/Z)^{1.16}}$$\times100$
  between the given pair is about 15\% which is same as in Ref. \cite{pak97,gaum10}. Based on the results
  of  E$_{bal}$ for isotopic pairs we concluded in Ref. \cite{gaum210} that mass dependence effects seem to
  dominate the isospin effects (consisting of isospin-dependent cross section, symmetry energy, surface
  properties). In our recent results \cite{sood10} we find that collective
  flow for isotopic pairs with large difference between N/Z is sensitive to the symmetry energy. Coulomb repulsion will be same for isotopic pair throughout the mass range.
   The comparison
  of  E$_{bal}$ for isotopic pairs gave us the hint that the Coulomb repulsion could
  be dominant in isospin effects in collective flow as well as its disappearance \cite{li96,pak97,gaum10}.
  Therefore Gautam and Sood \cite{gaum210} studied the isospin effects on the E$_{bal}$ throughout the
mass range 48-350 for two sets of isobaric systems with N/Z = 1.0 and 1.4 at semi central colliding geometry. These results
showed that the difference between the E$_{bal}$ for two isobaric
 systems is mainly due to the Coulomb repulsion. It was also shown that
  Coulomb repulsion dominates over symmetry energy.
These findings also indicated towards the dominance of the Coulomb
 repulsion in larger magnitude of isospin effects in
 E$_{bal}$ at peripheral collisions. Here we aim to extend the study over
  full range of colliding geometry varying from central to
 peripheral ones. Section 2
describes the model in brief. Section 3 explains the results and
discussion and Sec. 4 summarizes the results.
\section{The model}
 The present study is carried out within the framework of
IQMD model \cite{hart98}. The IQMD model treats
different charge states of nucleons, deltas, and pions explicitly,
as inherited from the Vlasov-Uehling-Uhlenbeck (VUU) model. The IQMD model has been used successfully for the
analysis of a large number of observables from low to relativistic
energies. The isospin degree of freedom enters into the
calculations via symmetry potential, cross sections, and Coulomb
interaction.
 \par
 In this model, baryons are represented by Gaussian-shaped density distributions
\begin{equation}
f_{i}(\vec{r},\vec{p},t) =
\frac{1}{\pi^{2}\hbar^{2}}\exp(-[\vec{r}-\vec{r_{i}}(t)]^{2}\frac{1}{2L})
\times \exp(-[\vec{p}- \vec{p_{i}}(t)]^{2}\frac{2L}{\hbar^{2}})
 \end{equation}
 Nucleons are initialized in a sphere with radius R = 1.12 A$^{1/3}$ fm, in accordance with liquid-drop model.
 Each nucleon occupies a volume of \emph{h$^{3}$}, so that phase space is uniformly filled.
 The initial momenta are randomly chosen between 0 and Fermi momentum ($\vec{p}$$_{F}$).
 The nucleons of the target and projectile interact by two- and three-body Skyrme forces, Yukawa potential, Coulomb interactions,
  and momentum-dependent interactions. In addition to the use of explicit charge states of all baryons and mesons, a symmetry potential between protons and neutrons
 corresponding to the Bethe-Weizsacker mass formula has been included. The hadrons propagate using Hamilton equations of motion:
\begin {eqnarray}
\frac{d\vec{{r_{i}}}}{dt} = \frac{d\langle H
\rangle}{d\vec{p_{i}}};& & \frac{d\vec{p_{i}}}{dt} = -
\frac{d\langle H \rangle}{d\vec{r_{i}}}
\end {eqnarray}
 with
\begin {eqnarray}
\langle H\rangle& =&\langle T\rangle+\langle V \rangle
\nonumber\\
& =& \sum_{i}\frac{p^{2}_{i}}{2m_{i}} + \sum_{i}\sum_{j>i}\int
f_{i}(\vec{r},\vec{p},t)V^{ij}(\vec{r}~',\vec{r})
 \nonumber\\
& & \times f_{j}(\vec{r}~',\vec{p}~',t) d\vec{r}~ d\vec{r}~'~
d\vec{p}~ d\vec{p}~'.
\end {eqnarray}
 The baryon potential\emph{ V$^{ij}$}, in the above relation, reads as
 \begin {eqnarray}
  \nonumber V^{ij}(\vec{r}~'-\vec{r})& =&V^{ij}_{Skyrme} + V^{ij}_{Yukawa} +
  V^{ij}_{Coul} + V^{ij}_{mdi} + V^{ij}_{sym}
    \nonumber\\
   & =& [t_{1}\delta(\vec{r}~'-\vec{r})+t_{2}\delta(\vec{r}~'-\vec{r})\rho^{\gamma-1}(\frac{\vec{r}~'+\vec{r}}{2})]
   \nonumber\\
   &  & +t_{3}\frac{\exp(|(\vec{r}~'-\vec{r})|/\mu)}{(|(\vec{r}~'-\vec{r})|/\mu)}+
    \frac{Z_{i}Z_{j}e^{2}}{|(\vec{r}~'-\vec{r})|}
   \nonumber \\
   &  & +t_{4}\ln^{2}[t_{5}(\vec{p}~'-\vec{p})^{2} +
    1]\delta(\vec{r}~'-\vec{r})
    \nonumber\\
   &  & +t_{6}\frac{1}{\varrho_{0}}T_{3i}T_{3j}\delta(\vec{r_{i}}~'-\vec{r_{j}}).
 \end {eqnarray}
Here \emph{Z$_{i}$} and \emph{Z$_{j}$} denote the charges of
\emph{ith} and \emph{jth} baryon, and \emph{T$_{3i}$} and
\emph{T$_{3j}$} are their respective \emph{T$_{3}$} components
(i.e., $1/2$ for protons and $-1/2$ for neutrons). The
parameters\emph{ $\mu$} and \emph{t$_{1}$,....,t$_{6}$} are
adjusted to the real part of the nucleonic optical potential.
 For the density dependence of  the nucleon optical potential, standard Skyrme type parametrization is employed.
 The momentum-dependence \emph{V$_{mdi}^{ij}$} of the nn interactions, which may optionally be used in IQMD, is fitted to the experimental data
 in the real part of the nucleon optical potential.
  It is worth mentioning
that the Gaussian width which describes the interaction range
of the particle depends on the mass of the system
in IQMD, since each nucleus shows maximum stability for a particular width as shown in Ref. \cite{hart98,gaum10} .
For eg, width of Ca is 4.16 fm$^{2}$ and for Au is 8.33 fm$^{2}$ and Gaussian width varies
between this mass range.

 \section{Results and discussion}
  For the present study, we simulate several thousands events of each
  reaction at incident energies around E$_{bal}$ in small steps of 10 MeV/nucleon.
   In particular, we simulate the reactions
$^{24}$Mg+$^{24}$Mg, $^{58}$Cu+$^{58}$Cu, $^{72}$Kr+$^{72}$Kr, $^{96}$Cd+$^{96}$Cd,
$^{120}$Nd+$^{120}$Nd, $^{135}$Ho+$^{135}$Ho, having N/Z = 1.0 and
reactions $^{24}$Ne+$^{24}$Ne, $^{58}$Cr+$^{58}$Cr, $^{72}$Zn+$^{72}$Zn,
$^{96}$Zr+$^{96}$Zr, $^{120}$Sn+$^{120}$Sn, and
$^{135}$Ba+$^{135}$Ba, having N/Z = 1.4, respectively,
 in the whole range of colliding geometry. The
colliding geometry is divided into four impact parameter bins of
0.15 $<$ $\hat{b}$ $<$ 0.25 (BIN 1),
 0.35 $<$ $\hat{b}$ $< $0.45 (BIN 2), 0.55 $<$ $\hat{b}$ $<$ 0.65 (BIN 3),
and 0.75 $<$ $\hat{b}$ $<$ 0.85 (BIN 4), where $\hat{b}$ =
b/b$_{max}$. Here N/Z is changed by keeping the mass fixed. We use
anisotropic standard isospin- and energy-dependent nn cross
section $\sigma$ = 0.8 $\sigma$$_{NN}$$^{free}$. The details about
the elastic and inelastic cross sections for proton-proton and
proton-neutron collisions can be found in Ref. \cite{hart98}.
The cross sections for neutron-neutron collisions are assumed to
be equal to the proton-proton cross sections.
\par
We also use soft equation of state along with momentum-dependent
intercations (MDI). The results with the above choice of equation of state and
cross section were in good agreement with the data
\cite{gaum10,gaum210}. The reactions are followed until the
transverse flow saturates. The saturation time varies form 100
fm/c for lighter masses to 300 fm/c for heavier masses. For transverse flow, we use the
quantity "\textit{directed transverse momentum $\langle p_{x}^{dir}\rangle$}" which is
defined as \cite{sood04,leh96}
\begin {equation}
\langle{p_{x}^{dir}}\rangle = \frac{1} {A}\sum_{i=1}^{A}{sign\{
{y(i)}\} p_{x}(i)},
\end {equation}
where $y(i)$ is the rapidity and $p_{x}$(i) is the momentum of
$i^{th}$ particle. The rapidity is defined as
\begin {equation}
Y(i)= \frac{1}{2}\ln\frac{{\textbf{{E}}}(i)+{{\textbf{p}}}_{z}(i)}
{{{\textbf{E}}}(i)-{{\textbf{p}}}_{z}(i)},
\end {equation}
where ${\textbf{E}}(i)$ and ${\textbf{p}_{z}}(i)$ are,
respectively, the energy and longitudinal momentum of $i^{th}$
particle. In this definition, all the rapidity bins are taken into
account. A straight line interpolation is used to calculate E$_{bal}$.
\begin{table}
\caption{The values of $\tau_{1}$ and $\tau_{1.4}$ for BIN 1 to BIN 4 for calculations with Coulomb potential
with and without A=48.}
\begin{center}
\begin{tabular}{|c|c|c|c|c|} \hline
 %& \multicolumn{8}{c|}{}\\%\cline{2-8}
&\multicolumn{2}{c|}{$\tau_{1}$}&\multicolumn{2}{c|}{$\tau_{1.4}$}\\\cline{2-5}
%Case&\multicolumn{2}{c|}{$Hard^{55}$}&\multicolumn{2}{c|}{$HMD^{55}$}&\multicolumn{2}{c|}{$Hard^{55}$}\\\cline{2-7}
b/b$_{max}$&  With A=48 &Without A=48  &With A=48  &Without A=48 \\\cline{1-5}
        BIN 1    &-0.36$\pm$ 0.01  &-0.38$\pm$ 0.03  &-0.33$\pm$ 0.01  &-0.31$\pm$ 0.02 \\\cline{1-5}
 \hline BIN 2    &-0.50$\pm$ 0.01  &-0.54$\pm$ 0.03  &-0.45$\pm$ 0.01  &-0.48$\pm$ 0.01 \\\cline{1-5}
 \hline BIN 3    &-0.70$\pm$ 0.03  &-0.83$\pm$ 0.07  &-0.56$\pm$ 0.02  &-0.57$\pm$ 0.07 \\\cline{1-5}
 \hline BIN 4    &-0.93$\pm$ 0.08  &-1.14$\pm$ 0.25  &-0.70$\pm$ 0.01  &-0.70$\pm$ 0.03 \\\cline{1-5}
\end{tabular}
\end{center}
\end{table}

\begin{figure}[!t] \centering
\vskip 0.5cm
\includegraphics[angle=0,width=13cm]{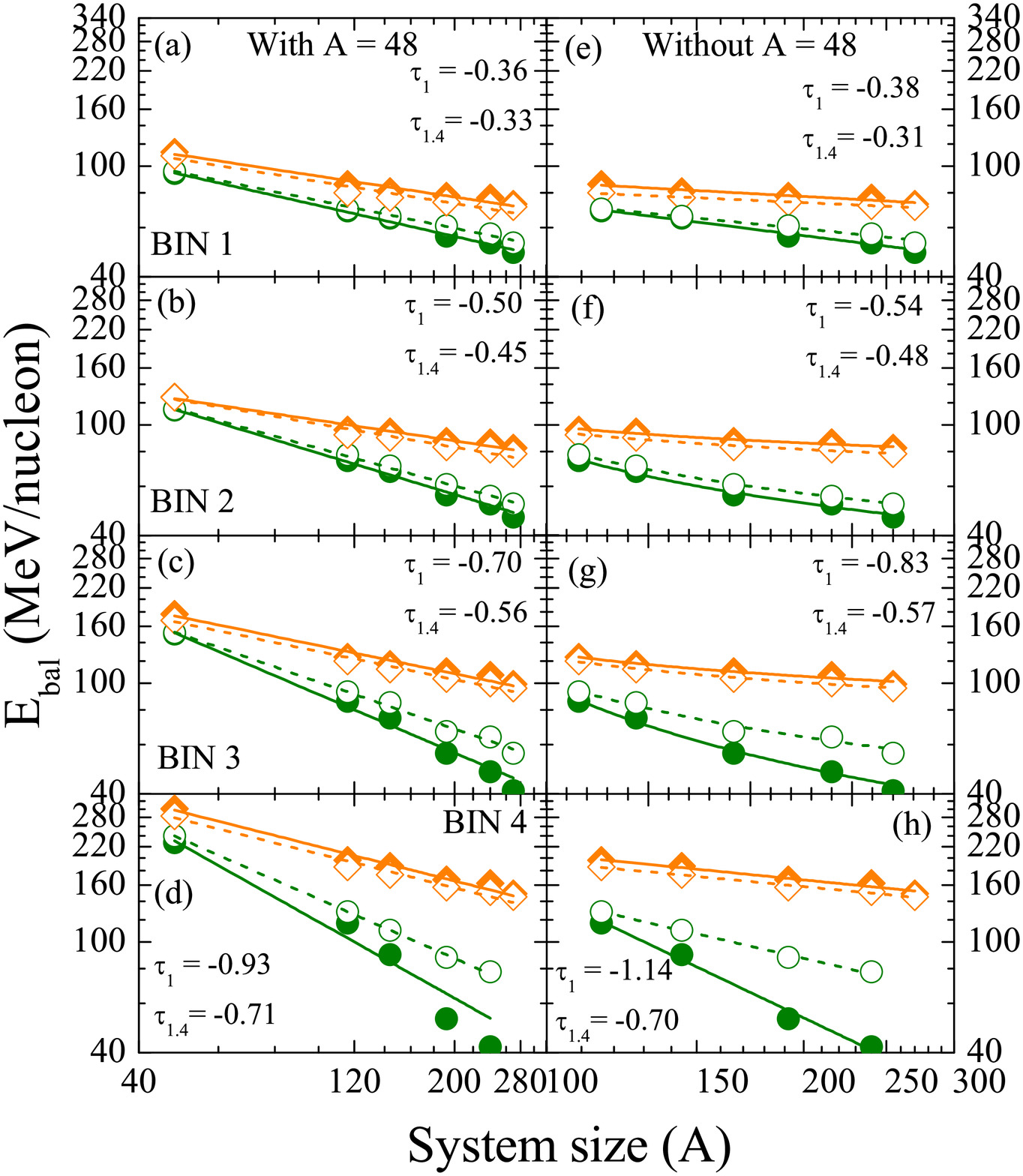}
\vskip 0.5cm \caption{(Color online) Left (right) panels
displays E$_{bal}$ as a function of
combined mass of system for different impact parameter bins with (without) A = 48. Solid
(open) symbols are for systems having N/Z = 1.0 (1.4). Circles (diamonds) are for calculations with full (reduced) Coulomb.
 Lines are power law fit $\varpropto
A^{\tau}$. $\tau$ values without errors for full Coulomb calculations are displayed in figure. The detailed values of $\tau$ are given in the table 1 and 2.}\label{fig1}
\end{figure}

\begin{figure}[!t] \centering
\vskip 0.5cm
\includegraphics[angle=0,width=10cm]{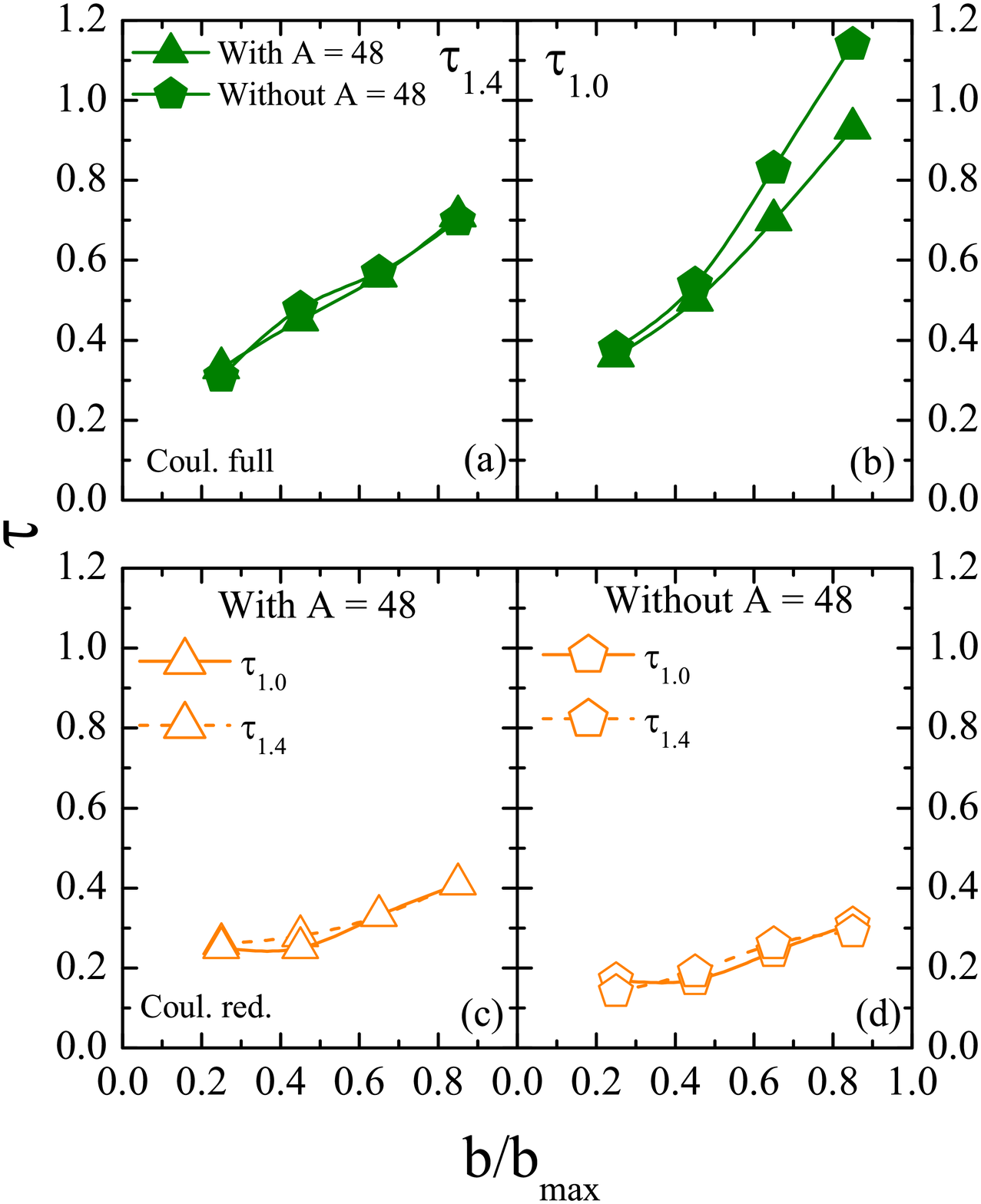}
\vskip 0.5cm \caption{(Color online) Upper (lower) panels display $\tau$ as a function of reduced impact
parameter for full (reduced) Coulomb. Values are plotted at upper limit of each bin. Symbols are explained in text. Lines are
only to guide the eye.}\label{fig3}
\end{figure}

Figure 1
displays the mass dependence of E$_{bal}$ for four impact parameter bins. The solid (open) green circles indicate
 E$_{bal}$ for systems with lower (higher) N/Z. Lines are power law fit $\varpropto$ A$^{\tau}$. Left (right) panels are for mass dependence of E$_{bal}$ when
 we include (exclude) A = 48. First we discuss the left panels for all the four bins. E$_{bal}$ follows a power law behavior
$\varpropto$ A$^{\tau}$ for both N/Z = 1 and 1.4 ($\tau$ being labeled as $\tau_{1.0}$ and $\tau_{1.4}$ for systems having
N/Z = 1 and 1.4, respectively) at all colliding geometries. There are small deviations from power law behavior
for heavy mass systems with N/Z = 1 at peripheral colliding geometry. Isospin effects are clearly visible for all the
  four bins as neutron-rich system has higher E$_{bal}$ throughout the mass range in agreement with the previous
  studies \cite{pak97,li96,gaum10}.

  \begin{table}
\caption{Same as table 1 but for calculations with Coulomb potential reduced by a factor of 100.}
\begin{center}
\begin{tabular}{|c|c|c|c|c|} \hline
 %& \multicolumn{8}{c|}{}\\%\cline{2-8}
&\multicolumn{2}{c|}{$\tau_{1}$}&\multicolumn{2}{c|}{$\tau_{1.4}$}\\\cline{2-5}
%Case&\multicolumn{2}{c|}{$Hard^{55}$}&\multicolumn{2}{c|}{$HMD^{55}$}&\multicolumn{2}{c|}{$Hard^{55}$}\\\cline{2-7}
b/b$_{max}$&  With A=48 &Without A=48  &With A=48  &Without A=48 \\\cline{1-5}
        BIN 1    &-0.25$\pm$ 0.02  &-0.17$\pm$ 0.02  &-0.26$\pm$ 0.03  &-0.14$\pm$ 0.01 \\\cline{1-5}
 \hline BIN 2    &-0.25$\pm$ 0.02  &-0.17$\pm$ 0.02  &-0.28$\pm$ 0.02  &-0.19$\pm$ 0.02 \\\cline{1-5}
 \hline BIN 3    &-0.33$\pm$ 0.02  &-0.24$\pm$ 0.03  &-0.33$\pm$ 0.02  &-0.26$\pm$ 0.01 \\\cline{1-5}
 \hline BIN 4    &-0.41$\pm$ 0.02  &-0.31$\pm$ 0.03  &-0.41$\pm$ 0.02  &-0.29$\pm$ 0.02 \\\cline{1-5}
\hline
\end{tabular}
\end{center}
\end{table}

\begin{figure}[!t] \centering
\vskip 0.5cm
\includegraphics[angle=0,width=10cm]{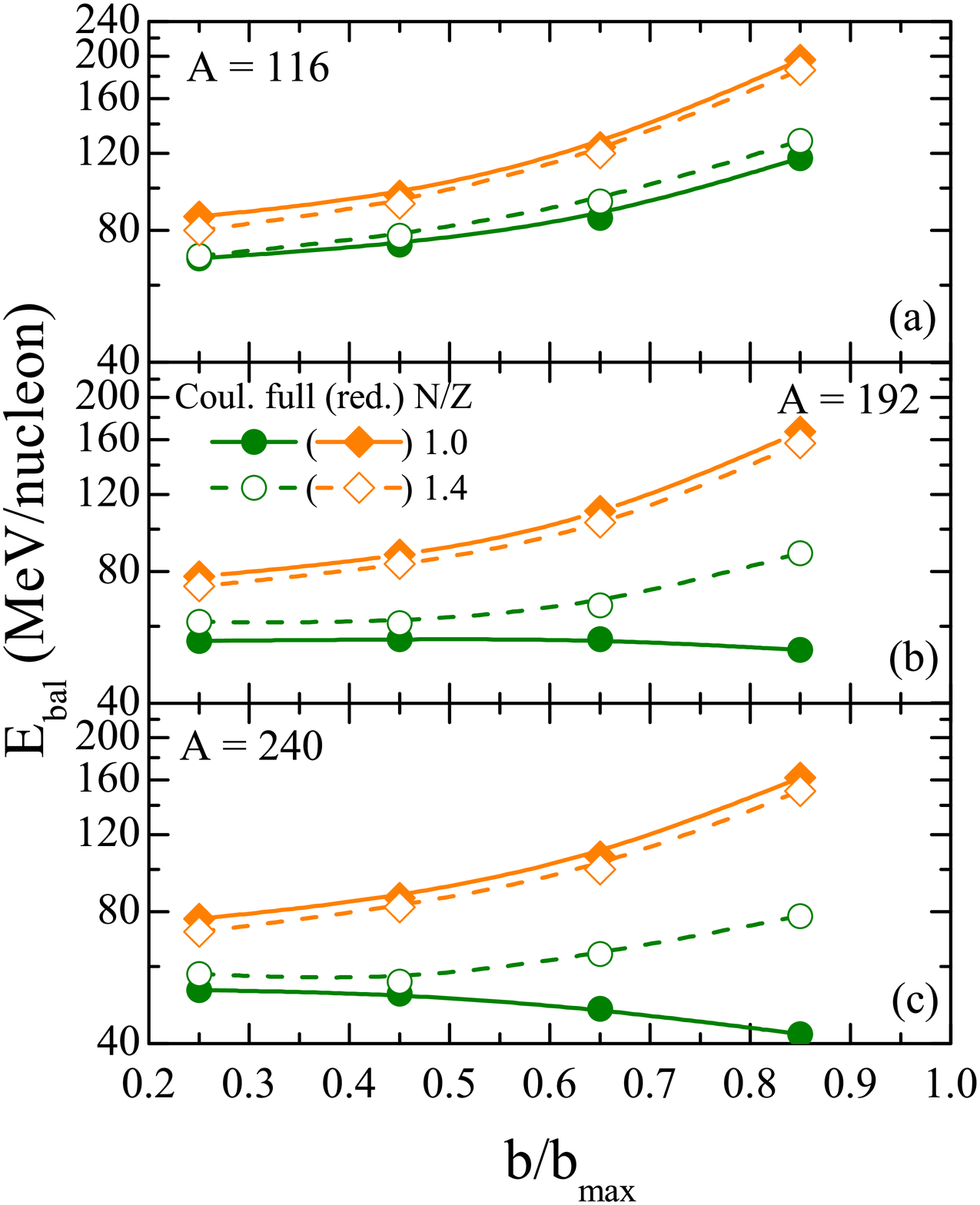}
\vskip 0.5cm \caption{(Color online) E$_{bal}$ as a function of impact parameter for
different system masses. Various symbols have same meaning as in Fig.1. Lines are only to guide the eye.}\label{fig4}
\end{figure}
The magnitude of isospin effects increases with increase in mass of the system at all colliding geometries. The effect
is much more pronounced at larger colliding geometries. One can see that the difference between
$\tau_{1.0}$ and $\tau_{1.4}$ increases with increase in the impact parameter (green circles). In Ref. \cite{gaum210}, Gautam and Sood studied the isospin
effects on the mass dependence of E$_{bal}$ for BIN 2. There they reduced the Coulomb potential by a factor of 100 and showed that the Coulomb
repulsion plays dominant role over symmetry energy in isospin effects on E$_{bal}$ as well as its mass dependence
 at semi central colliding geometry (BIN 2). Since here we plan to extend
that study over a full range of colliding geometry, so here also we reduce the Coulomb potential by a factor
of 100 and calculate the E$_{bal}$ throughout the mass range at all colliding geometries. Solid (open) diamonds
 represent E$_{bal}$ calculated with reduced Coulomb for systems with lower (higher) neutron content.
  Lines are power law fit $\varpropto$ A$^{\tau}$. Interestingly, we find that:\\
(a) the magnitude of isospin effects (difference in E$_{bal}$ for a given pair)
is now nearly same throughout the mass range which indicates that the effect of symmetry energy is uniform
throughout the mass range. This is true for all the colliding geometries. This is supported by
 Ref. \cite{sood204} where Sood and Puri studied the average density as a function of mass of the system
 (throughout the periodic table) at incident energies equal to E$_{bal}$ for each
given mass. There they found that although both E$_{bal}$ and average density follows a power law behavior $\varpropto$ A$^{\tau}$, E$_{bal}$ decreases more sharply with the
 combined mass of the system (with $\tau$ = -0.42), whereas the average
density (calculated at incident energy equal to E$_{bal}$) is almost independent of the mass of the system with $\tau$ = -0.05.
It is worth mentioning here that the trend will be different at fixed incident energy in which case density increases with increase
 in the mass of the system \cite{blatt,khoa}. We also note that the power law with reduced Coulomb is now absolute for N/Z = 1 (solid diamonds fig. 1 (d)).
 This indicates that the deviations from power law behavior for heavy mass systems (with N/Z = 1) are due to the
 dominance of Coulomb repulsion.\\
 (b) one can also see that the enhancement in E$_{bal}$ (by reducing
Coulomb) is more in heavier
 systems as compared to lighter systems for
 all colliding geometries. The effect is more pronounced at higher colliding geometries. \\
 (c) throughout the mass range at all
  colliding geometries, the neutron-rich systems
 have a decreased E$_{bal}$ as compared to neutron-deficient systems when we reduce the Coulomb. This trend is quite the opposite to the one
 which we have when we have full Coulomb. This (as explained in Ref. \cite{gaum210} also) is due to the fact that the reduced
Coulomb repulsion leads to higher E$_{bal}$. As a result, the
density achieved during the course of the reaction will be more
due to which the impact of the repulsive symmetry energy will be
more in neutron-rich systems, which in turn leads to a decreased
E$_{bal}$ for neutron-rich systems.
\par
Now to discuss how the value of $\tau$ changes if we exclude lighter systems (right panels in fig. 1), we display in fig. 2 the variation
of $\tau$ as a function of impact parameter.
 Solid (open) symbols
are for full (reduced) Coulomb. Triangles (pentagons) represent $\tau$ with (without) A = 48. For full Coulomb (upper left panel), $\tau_{1.4}$
increases with increase in impact parameter but the increase is independent of inclusion/exclusion of lighter mass. On the other
hand, $\tau_{1}$ increases drastically with impact parameter (upper right panel). Moreover, the increase in $\tau_{1}$ (with impact parameter)
 is more sharp when we take into account only heavier masses (pentagons) as compared to when lighter mass systems are also included (triangles).
Thus as we have discussed in Ref. \cite{gaum210} also that if we take into account only
 heavier systems, the value of $\tau$ is more as compared to if we include lighter systems as well. Here we see enhancement
 of this effect with increase in colliding geometry showing the increased role of Coulomb repulsion in mass dependence of
E$_{bal}$ at high impact parameter.
    In lower left panel, for reduced Coulomb, we see the values of $\tau_{1}$ and $\tau_{1.4}$
 increases much less sharply with increase in colliding geometry and this increase is almost independent of N/Z of the system
 which shows that the effect of isospin-dependent cross section and
 symmetry energy on the mass dependence of E$_{bal}$ is independent of N/Z throughout the colliding geometry. The same is true if we consider
 only heavier systems as well (lower right panel). As we have seen in fig. 1, the effect of symmetry energy is uniform throughout the mass range
 at all colliding geometries, this means that the effect of cross section must also be nearly same through the mass range and colliding geometry.
 We will come to this point later. It is worth mentioning that since IQMD reproduces the data nicely \cite{gaum10} so the experimental values of tau are not expected to differ much from the values of tau given in table 1.

 \par
\begin{figure}[!t] \centering
\vskip 0.5cm
\includegraphics[angle=0,width=10cm]{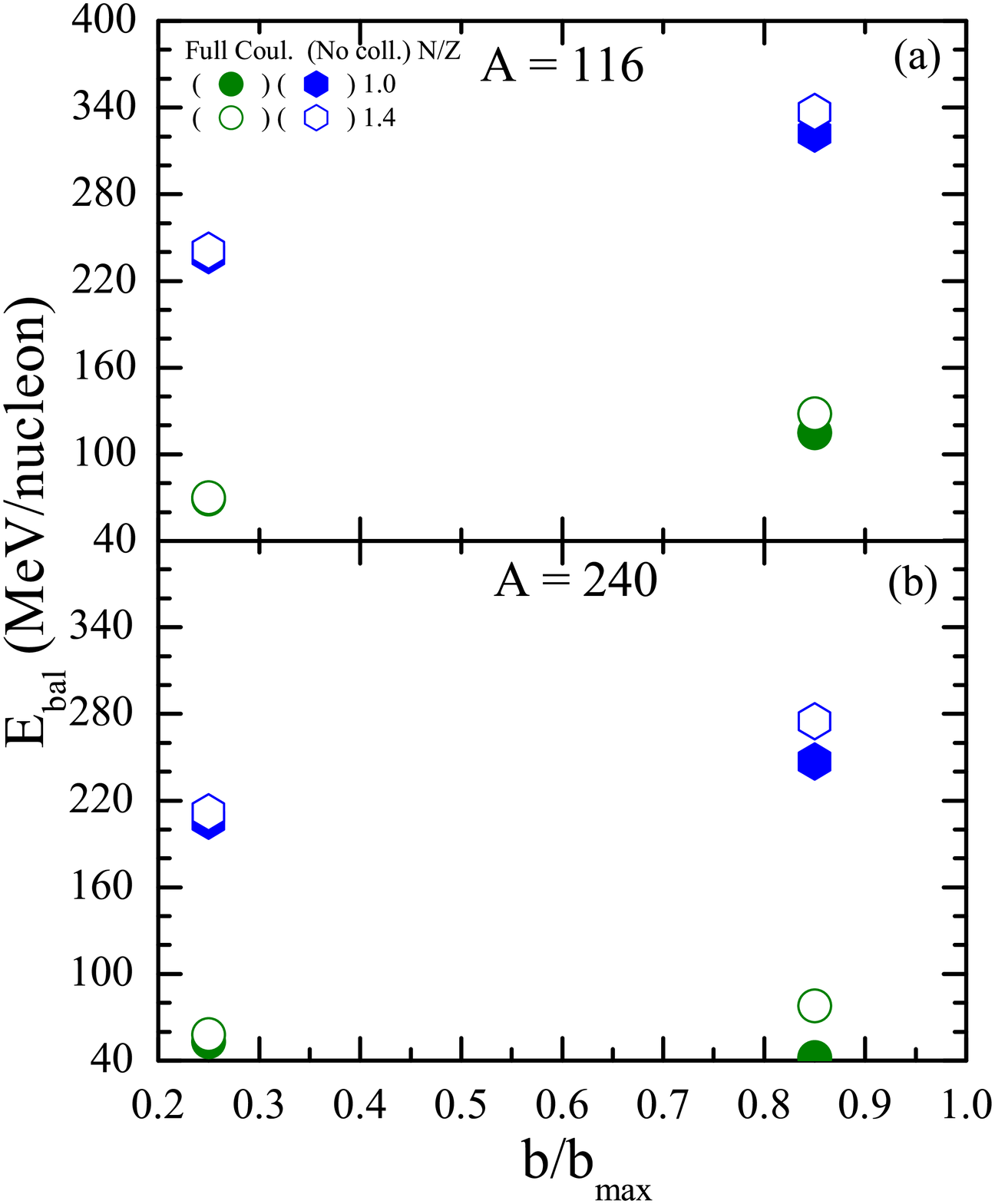}
\vskip 0.5cm \caption{(Color online) E$_{bal}$ at central and peripheral colliiding geometries
for A = 116 (upper panel) and A = 240 (lower panel) with no nucleon-nucleon collisions (Hexagons). Circles
represent the calculations with collisions.}\label{fig4}
\end{figure}

\begin{figure}[!t] \centering
\vskip 0.5cm
\includegraphics[angle=0,width=10cm]{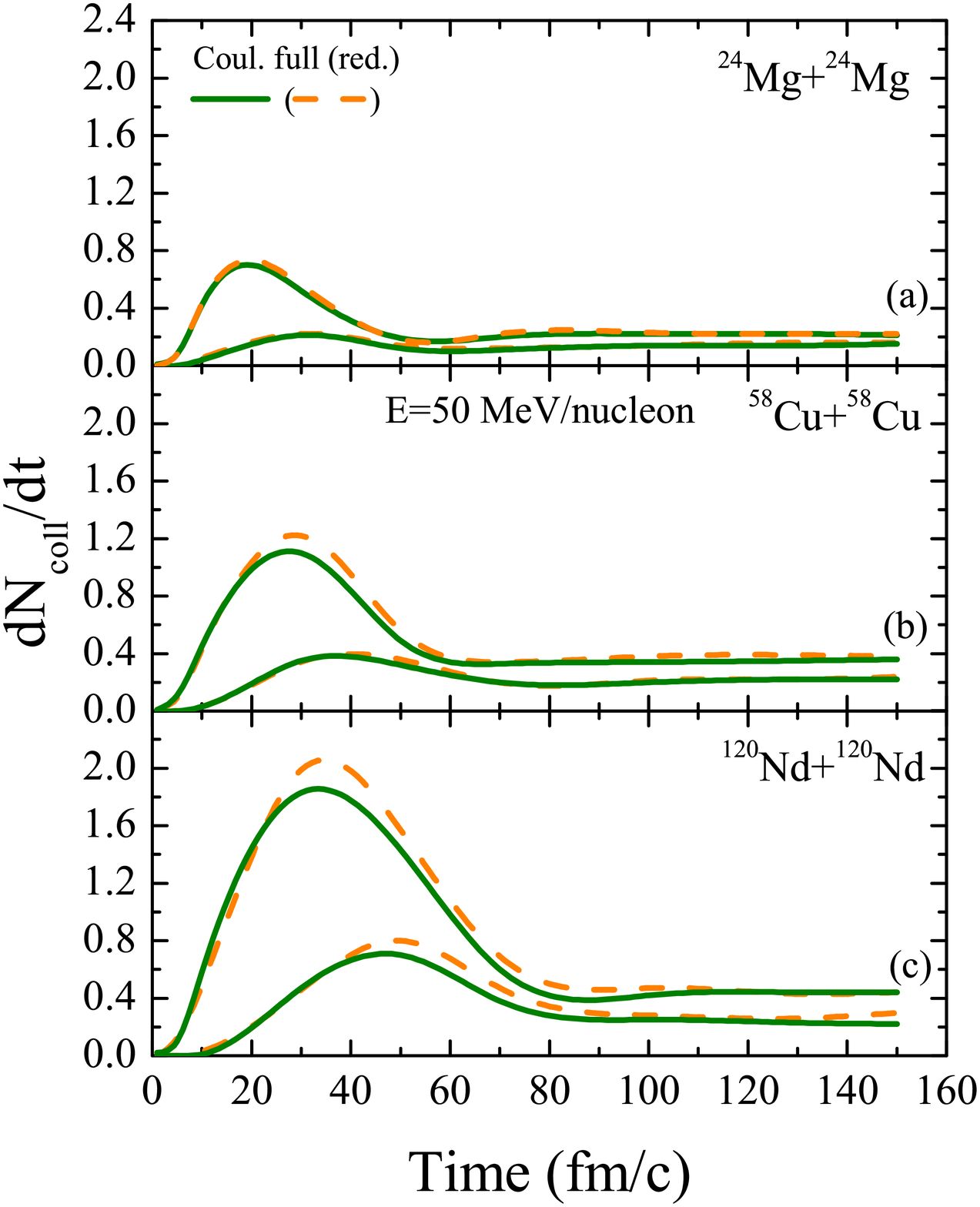}
\vskip 0.5cm \caption{(Color online) The time evolution of collision rate ($\frac{dN_{coll}}{dt}$) for various
system masses at 50 MeV/nucleon for BIN 1 and BIN 4. Various lines are explained in the text .}\label{fig4}
\end{figure}

 In fig. 3a, 3b, and 3c, we display E$_{bal}$ as a function of $\hat{b}$ for masses 116, 192, and 240, respectively, for both full
 and reduced Coulomb. Symbols have the same meaning as in fig. 1. For full Coulomb (green circles), for all the masses at all colliding
 geometries, system with higher N/Z has larger E$_{bal}$ in agreement with previous studies \cite{li96,pak97,mag00}. Moreover,
  the difference between E$_{bal}$ for a given mass
 pair, increases with increase in colliding geometry. This is more clearly visible in heavier masses. Also for N/Z = 1.4, E$_{bal}$ increases with increase in
 impact parameter in agreement with Ref. \cite{mag00}. This is due to the decreased participant zone in
 peripheral collisions which decreases the amount of repulsive nn collisions and therefore
  higher energy is required to overcome the attractive nuclear force. This effect is less pronounced in heavier
systems since even at peripheral geometry significant number of nn collisions will occur. Moreover,
 the effect of stronger Coulomb repulsion in heavier systems will increase with colliding
 geometry since it will push more number of nucleons in the transverse direction away from participant zone.
  However for N/Z = 1, increase in E$_{bal}$ with impact parameter is true only for lighter mass system such as A = 116.
 For heavier masses E$_{bal}$ infact
 begins to decrease with increase in impact parameter in contrast to the previous studies \cite{li96,pak97,gaum10,mag00}.
 However, when we reduce the Coulomb (by a factor of 100 (diamonds)),
 we find that:
 \par
  (i) Neutron-rich systems have a decreased E$_{bal}$ as compared to neutron-deficient systems as mentioned
  previously also. This is true at all the
colliding geometries throughout the mass range. This clearly shows the dominance of Coulomb repulsion over symmetry energy in isospin effects throughout the mass range at all colliding geometries.
\par
(ii) The difference between E$_{bal}$ for systems with different N/Z remains almost constant as a
function of colliding geometry which indicates that the effect of symmetry energy is uniform throughout the range of $\hat{b}$ as well. This also shows that
the large differences in E$_{bal}$ values for a given isobaric pair are due to the Coulomb repulsions.
 \par
(iii) In heavier systems, at high colliding geometry, the increase in E$_{bal}$  is more in systems with N/Z =1 compared to N/Z = 1.4 which shows the
much dominant role of Coulomb repulsion at high colliding geometry.
\par
To see the relative contribution of Coulomb repulsion and cross section
in lighter and heavier systems, we switch off the collision term and calculate the E$_{bal}$ for A = 116 and 240 at two extreme bins. The results are displayed
in fig. 4. Hexagons represent the calculations without collisions. Other symbols have same meaning as in fig. 1. We find that at a given impact parameter
E$_{bal}$ increases by large magnitude for both systems which shows the importance of collisions. The increase is of the same order for
both the masses at both impact parameter bins indicating the same role of cross section for both lighter and heavier masses as we have
expected in discussion of fig. 2. This is supported by
Ref. \cite{sood204,blatt}, which shows that since mean field is independent of the mass of the system, so one needs the same amount of collisions to counter
balance the mean field in both lighter and heavier masses. Moreover, the same order of increase of E$_{bal}$ at central and peripheral colliding geometry
(when we switch off the collisions) indicates the importance of collisions at high colliding geometry as well. The effect that
E$_{bal}$ decreases with increase in impact parameter (due to dominance of Coulomb) for heavy mass systems with N/Z = 1 (fig. 3 lower panel)
 does not appear here for lighter and
heavier masses. Therefore in fig. 3, the reduced Coulomb allows one to examine the balance of nn collisions and mean field while in fig. 4
the removal of nn collisions allows one to examine the balance of Coulomb and mean field.
\par
As we have seen in fig. 4 that E$_{bal}$ is much higher than the actual E$_{bal}$ when we switch off
the cross section, so to explore whether Coulomb repulsion affects the collisions in the E$_{bal}$ domain, we display in fig. 5 collision rate
 $\frac{dN_{coll}}{dt}$ for A = 48 (upper panel), 116 (middle panel), and 240 (bottom panel) at incident energy of 50 MeV/nucleon. Solid (dashed) lines represent
 Coulomb full (reduced) calculations. Higher (lower) peaks represent results for central (peripheral) impact parameter.
 We find that Coulomb decreases the collision rate in medium and heavier mass systems for central bins whereas for peripheral bins, the effect
 of Coulomb on collisions is only for heavier masses (lower panel). Comparing
 top and bottom panel we see that there are still significant number of collisions at peripheral geometry.
\section{Summary}
We have studied the isospin effects in the disappearance of flow
as well as its mass dependence throughout the mass range 48-270 for two
sets of isobaric systems with N/Z = 1 and 1.4 in the whole range
of colliding geometry. Our results clearly demonstrate the
dominance
 of Coulomb repulsion in isospin effects on E$_{bal}$ as well as its
mass dependence throughout the range of colliding geometry. The above study also shows that the effect of symmetry
energy on E$_{bal}$ and cross section is uniform throughout the
mass range and colliding geometry. We have also presented the counter balancing of nn collisions and mean field by reducing the Coulomb and the counter
balancing of Coulomb and mean filed by removing the nn collisions.
\par
This work has been supported by a grant from Indo-French Centre For The Promotion Of Advanced Research (IFCPAR) under project no. 4104-1.

%%%%%%%%%%%%%%%%%%%%%%%%%%%%%%%%%%
\end{document}